\documentclass[prb,twocolumn,superscriptaddress,nofootinbib,noshowpacs,preprintnumbers,longbibliography,floatfix]{revtex4-1}

\usepackage{amsthm}
\usepackage{mathtools}
\usepackage{physics}
\usepackage{xcolor}
\usepackage{graphicx}
\usepackage[left=23mm,right=13mm,top=35mm,columnsep=15pt]{geometry} 
\usepackage{adjustbox}
\usepackage{placeins}
\usepackage[T1]{fontenc}

\usepackage[english]{babel}
\usepackage[utf8]{inputenc}
\usepackage{csquotes} 

\usepackage[pdftex, pdftitle={Article}, pdfauthor={Author}]{hyperref} 

\usepackage{bm}
\renewcommand{\vec}{\boldsymbol}

\newcommand{\refEq}[1]{Eq.~(\ref{#1})}
\newcommand{\refFig}[1]{Fig.~\ref{#1}}

\begin{document}
\title{The quantum skyrmion Hall effect in \texorpdfstring{$f$}{f} electron systems}

\author{Robert Peters}
    \email[Correspondence email address: ]{peters.robert.7n@kyoto-u.ac.jp}
    \affiliation{Department of Physics, Kyoto University, Kyoto 606-8502, Kyoto, Japan}
\author{Jannis Neuhaus-Steinmetz}
    \affiliation{Department of Physics, University of Hamburg, 20355 Hamburg, Germany}
    \affiliation{The Hamburg Centre for Ultrafast Imaging, Luruper Chaussee 149, 22761 Hamburg, Germany}
\author{Thore Posske}
    \email[Correspondence email address: ]{thore.posske@uni-hamburg.de}
    \affiliation{I. Institute for Theoretical Physics, Universit{\"a}t Hamburg, Notkestra{\ss}e 9, 22607 Hamburg, Germany}
    \affiliation{The Hamburg Centre for Ultrafast Imaging, Luruper Chaussee 149, 22761 Hamburg, Germany}


\begin{abstract}
The flow of electric current through a two-dimensional material in a magnetic field gives rise to the family of Hall effects. 
The quantum versions of these effects accommodate robust electronic edge channels and fractional charges.
Recently, the Hall effect of skyrmions, classical magnetic quasiparticles with a quantized topological charge, has been theoretically and experimentally reported, igniting ideas on a quantum version of this effect.
To this end, we perform dynamical mean field theory calculations on localized $f$ electrons coupled to itinerant $c$ electrons in the presence of spin-orbit interaction and a magnetic field.
Our calculations reveal localized nano quantum skyrmions that start moving transversally when a charge current in the itinerant electrons is applied.
The results show the time-transient build-up of the quantum skyrmion Hall effect, accompanied by an Edelstein effect and a magnetoelectric effect that rotate the spins.
This work motivates studies about the steady state of the quantum skyrmion Hall effect, looking for eventual quantum skyrmion edge channels and their transport properties.
\end{abstract}


\maketitle

\section{Introduction}

From fundamental physical processes to application-oriented information storage and processing, the stability of a physical effect is paramount. Some physical effects, especially quantum Hall effects, accommodate observables that are topologically protected, i.e., they are robust against a continuous deformation of selected parameters. Recently, experimental and theoretical studies have found topologically protected classical magnetic structures in thin films or effectively two-dimensional systems, which have been coined magnetic skyrmions \cite{Bogdanov1995,BogdanovHubert1999,Muhlbauer2009,YuOnose2010,Heinze2011}, connected to earlier ideas in particle physics \cite{Skyrme1962}.
The stability of these objects and the possibility of creating them by electrical currents or time-controlled magnetic boundary conditions \cite{Evershor-Sitte2017,Stier2017,Schaffer2020,Siegl2022Racetrack} promote the idea of using them in spintronics and as information carriers \cite{Parkin..Thomas2008MagneticDomainWallRacetrackMemory,Tomasello..Finocchio2014AStrategyForTheDesignOfSkyrmionRacetrackMemories,Fert2017}.
Furthermore, in sight of the ongoing miniaturization of magnetic skyrmions, they have also been proposed as ingredients in quantum computing  
\cite{Psaroudaki2021}. 

Classical magnetic skyrmions experience an additional drag transversal to the direction of an applied electric current, which leads to an accumulation of skyrmions at one side \cite{Thiele1973,Evershor-Sitte2011,Evershor-Sitte2012,Iwasaki2013a,Iwasaki2013b,Sampaio2013,Moon2022,Jiang2016DirectObservationSkyrmionHallEffect,Litzius2016}. 
The angle between the direction of motion and the direction of the current is called the Magnus angle of the skyrmion Hall effect.
The question arises if there is a quantum version of the skyrmion Hall effect and, if so, which characteristics of the electronic Hall effects transfer, including hypothetical skyrmion edge channels with their quantized conductance. A previous study treating the quantum skyrmion as a product state and calculating an effective action for the quantum skyrmion demonstrated the existence of the Magnus force \cite{PhysRevB.94.134415}.
Yet, the challenge in describing general quantum skyrmions comes with the large Hilbert spaces that need to be considered in two-dimensional spin systems carrying quantum skyrmions of a size of minimally $3\times 3$ spins \cite{Sotnikov2021,Lohani2019,Siegl2022}, which demand special theoretical techniques like density matrix renormalization group \cite{Haller2022} or artificial neural networks \cite{AshishPetersPosske2022} to investigate them numerically.

Another method to analyze quantum skyrmions
is the use of localized spins from interacting electrons, like $f$ electrons, to represent the skyrmions in a correlated electronic system \cite{PhysRevB.106.L140406}. Representing the skyrmions as electronic degrees of freedom has the advantage that we can treat considerably large quantum spin systems with established advanced numerical techniques for correlated electronic systems and that we can apply an electric current within the model without further assumptions. Interestingly, skyrmions in $f$-electron systems have been experimentally detected in EuAl$_4$, in which the skyrmions have been treated classically \cite{Nat_com13_1472}. Yet, the quantum nature of skyrmions in strongly correlated electronic systems is not well studied.
Ultimately, a quantum skyrmion Hall effect could have direct practical applications extenting the manifold of suggested technical applications of magnetic skyrmions \cite{Back2020SkyrmionRoadmap} to the quantum world. Moreover, fundamental questions about the topological nature of quantum skyrmions, which, strictly speaking, gets lost because of quantum spin slip processes \cite{KimTserkovnyak2016TopologicalEffectsOnQuantumPhaseSlipsInSuperfluidSpinTransport,Posske2019,Siegl2022,Vijayan2023}, could be answered when quantum skyrmions are connected to quantum Hall effects and their unambiguous topological origin. 

In this paper, we numerically study a square lattice of localized $f$ electrons that are coupled to two-dimensional itinerant conduction ($c$) electrons in the presence of spin-orbit coupling and a small magnetic field perpendicular to the two-dimensional plane. 
Using dynamical mean-field theory, we reliably identify regions in parameter space that host quantum nano skyrmions.
We subsequently study the effect of a current in the itinerant electrons on the quantum skyrmion in linear response theory and find a strong initial drag into the direction perpendicular to the current, marking the onset of the quantum skyrmion Hall effect.  The shift of the skyrmion is accompanied by an Edelstein effect and a magnetoelectric effect,\cite{EDELSTEIN1990233,PhysRevLett.99.226601,Chernyshov2009,PhysRevB.78.212405,PhysRevB.80.134403,PhysRevB.97.115128,Fiebig_2005} which leads to a rotation of the localized $f$-electron spins.
Our study stimulates further investigation of the quantum skyrmion Hall effect, especially its steady state, and possibly quantized skyrmion edge channels. 

The remainder of this paper is structured as follows: In Sec.~\ref{Sec:Model}, we introduce the model and the method. In Sec.~\ref{Sec:stability}, we analyze the stability and structure of the quantum skyrmions for different model parameters. This is followed by Sec.~\ref{Sec:SkyrmionHall}, where we demonstrate the onset of the skyrmion Hall effect using linear response theory. Finally, in Sec.~\ref{Sec:discussion}, we discuss our results and conclude the paper.

\section{Model and Method}
\label{Sec:Model}
Motivated by the discovery of magnetic skyrmions in Eu-compounds \cite{Nat_com13_1472}, including partially filled $f$ electrons, we focus here on magnetically ordered ground states and low-energy metastable states in $f$-electron systems  on a square lattice with a lattice constant of $a$ on the order of half a nanometer. 
In particular, we study the ground states of a noncentrosymmetric $f$-electron system described by a periodic Anderson model \cite{JPSJ.77.124711,PhysRevB.97.115128,PhysRevB.99.155141}.
It is important to note that we explicitly start with an electronic Hamiltonian instead of a quantum spin model. Thus, charge fluctuations and other effective interactions besides the effective Heisenberg and Dzyaloshinskii–Moriya (DM) interaction generally affect the ground state. Furthermore, due to the hybridization between the itinerant conduction ($c$) electrons and the $f$ electrons, the magnetic moments generated by the $f$ electrons are intrinsically coupled to the $c$ electrons. Such a coupling, which is necessary to observe skyrmion Hall and skyrmion drag effects, does hence not need to be inserted manually but is naturally included.

Our model Hamiltonian can be split into a single-particle part, $H_0$, and an interaction part, $H_U$. The single-particle Hamiltonian is
\begin{eqnarray}
\label{eqnHamiltonian}
H_0(\vec{k})&=& 
\left(t\left[\cos(k_x)+\cos(k_y)\right]+\left[\mu_c+\mu_f\right]/{2} \right) \vec{c}^\dagger_{\vec k}\vec{c}_{\vec k}
\nonumber \\
&+&
\left(t
    \left[\cos(k_x)+\cos(k_y)\right]
        +\left[\mu_c-\mu_f\right]/{2}
    \right)
\vec{c}^\dagger_{\vec k}\sigma^0 \tau^z\vec{c}_{\vec k}
\nonumber \\
&-& 2\alpha_c \sin(k_y)  \vec{c}^\dagger_{\vec k} \sigma^x \tau^x  \vec{c}_{\vec k} 
+2\alpha_c \sin(k_ x) \vec{c}^\dagger_{\vec k} \sigma^y \tau^x \vec{c}_{\vec k}
\nonumber \\
&+& V \vec{c}^\dagger_{\vec k} \sigma^0 \tau^x \vec{c}_{\vec k}
+ B \vec{c}^\dagger_{\vec k,\rho_1\tau_1} \sigma^z \tau^0 \vec{c}_{\vec k},
\end{eqnarray}
where 
$\vec{c}_{\vec{k}} = 
\left(
  c_{k_x,k_y,\uparrow},
  f_{k_x,k_y,\uparrow},
  c_{k_x,k_y,\downarrow},
  f_{k_x,k_y,\downarrow}
  \right)$
  is the spinor containing the momentum space annihilation operators of the itinerant electrons and $f$ electrons, respectively, corresponding to the real-space operators 
  $c_{i,j,\sigma}$ and $f_{i,j,\sigma}$ at site $\left(i,j\right)$ of a square lattice with spin $\sigma$. 
  The matrices $\sigma^\lambda = s^\lambda \otimes s^0$ and $\tau= s^0 \otimes s^\lambda$ denote the Pauli matrices on the spin and sublattice space, respectively, where $s$ are the bare Pauli matrices. 
The particle number operators are 
$n^c_{i,j,\sigma}=c^\dagger_{i,j,\sigma}c_{i,j,\sigma}$ and $n^f_{i,j,\sigma}=f^\dagger_{i,j,\sigma}f_{i,j,\sigma}$.
The strength of the nearest neighbor hopping of the $c$ electrons on the square lattice is denoted by $t$. Throughout this paper, we use $t$ as the unit of energy. $\mu_c$ and $\mu_f$ are local energies of the $c$ and $f$ electrons, respectively. $V$ is a local hybridization between the $c$ and $f$ electrons as commonly used in the periodic Anderson model. $B$ corresponds to a small magnetic field applied in the $z$ direction.
Finally, we include a spin-orbit coupling between the $c$ and $f$ electrons with hopping amplitude $\alpha_{c}$. This spin-orbit coupling corresponds to a Rashba-type spin-orbit interaction as present in noncentrosymmetric $f$-electron systems \cite{JPSJ.77.124711}. 
The interaction part of the Hamiltonian is
\begin{equation}
H_U=U\sum_{i,j} n^f_{i,j,\uparrow} n^f_{i,j,\downarrow},
\end{equation}
corresponding to a density-density interaction between $f$ electrons on the same lattice site.
The full Hamiltonian is 
\begin{equation}
    H=H_0+H_U.
\end{equation}
The calculations are performed on a finite lattice $L_x\times L_y=11\times 11$ with periodic boundary conditions.

To find the ground state of this quantum model, we use the real-space dynamical mean-field theory (RDMFT)\cite{RevModPhys.68.13,PhysRevB.59.2549,SciPostPhys.11.4.083,PhysRevB.92.075103,PhysRevB.89.155134}. RDMFT maps each atom of a unit cell (finite lattice) on its own quantum impurity model by calculating the local Green's function
\begin{equation}
    G_{n,m}(z)=\left( z-\tilde{h}_0-\Sigma(z)\right)^{-1}_{n,m},\label{EQ:localGreen}
\end{equation}
where $\tilde{h}_0$ is the single-particle matrix of the Fourier transform of $H_0$ in Eq.~(\ref{eqnHamiltonian}) into real space, i.e., $\tilde{H}_0 = \sum_{n,m} c^\dagger_n \tilde{h}_{n,m} c_m$. Here, $n$ and $m$ are super indices including the lattice positions, the $f$-$c$ sublattice, and the spin. Furthermore, $\Sigma(z)$ is a matrix including the local self-energies of each lattice site in the finite lattice, where, by the defining approximation of RDMFT, $\Sigma_{n,m}(z)$ vanishes when the spatial components of $n$ and $m$ differ.
Writing the local Green's function as 
\begin{equation}
    G_{n,m}=\left({z-\Delta_{n,m}(z)-\Sigma_{n,m}(z)}\right)^{-1},
\end{equation}
we can map each lattice site on a quantum impurity model, 
where $\Delta_{n,m}(z)$ is the local hybridization of the impurity model.
This hybridization function describes the environment for one lattice site created by the rest of the lattice.
Here, the self-energy differs for each lattice site, and hence this hybridization function is different for each lattice site.
Summarizing the numerical procedure, the local hybridization functions define quantum impurity models, which are solved to obtain the local self-energy for each lattice site. These updated self-energies are then used in Eq.~(\ref{EQ:localGreen}), which defines a self-consistency problem. To calculate the self-energy of each lattice site, we use the numerical renormalization group (NRG) \cite{RevModPhys.47.773,RevModPhys.80.395,PhysRevB.74.245114}, which can calculate accurate Green's functions and self-energies at low temperatures.

The magnetic properties of the periodic Anderson model without Rashba spin-orbit interaction are well understood within the DMFT approximation \cite{RevModPhys.68.13}. At half-filling, $\langle n^c_{i,j}\rangle=\langle n^f_{i,j}\rangle=1$, on a square lattice, the periodic Anderson model orders antiferromagnetically for weak hybridization strengths $V$ \cite{PhysRevB.92.075103}. For large hybridization strengths, the periodic Anderson model at half-filling becomes a Kondo insulator. On the other hand, when the number of $c$ electrons is small and the $f$ electrons are nearly half-filled, the system orders ferromagnetically \cite{PhysRevLett.108.086402}. 
This paper aims to study the existence and properties of magnetic skyrmions in a ferromagnetic periodic Anderson model, including  Rashba spin-orbit interaction. We thus look for parameters where the $f$ electrons are nearly half-filled, and the $c$-electron filling is about $\langle n^c\rangle\approx 0.2$.

An exhaustive search of the parameter space of the periodic Anderson model for stable quantum skyrmions in the ground state is numerically unfeasible.
In advance to our fully quantum-mechanical calculations, we therefore first identify candidate parameter regions where the ground state or low-energy metastable states accommodate magnetic skyrmions. We do so by mapping \refEq{eqnHamiltonian} to a classical Heisenberg spin model with nearest-neighbor coupling by integrating out the $c$ electrons using second-order perturbation theory, 
which obtains the RKKY spin-spin interactions \cite{RudermannKittel1954,Kasuya1956,Yosida1957}. 
We then use classical Monte Carlo methods to find the ground states of these spin models \cite{NeuhausSteinmetz2022}. In particular, we have varied in this procedure the local hybridization $V$, the spin-orbit coupling $\alpha_c$, and the $c$-electron level position, $\mu_c$. 
We subsequently transfer parameter configurations where we find a classical skyrmion to the quantum model and conduct RDMFT calculations to obtain the system's ground state. Here, the presence of magnetic skyrmions in the corresponding classical model generally is a good indicator for quantum skyrmions in the quantum model. 
Setting $U/t=6$ and $\mu_f/t=-3$, corresponding to half-filling of the $f$ electrons, we find quantum skyrmions in a ferromagnetic background for $V=t$, $\mu_c/t\approx 3.6$, and a finite spin-orbit coupling in combination with a magnetic field, in agreement with previous results on classical and quantum magnetic skyrmions \cite{Sotnikov2021,Lohani2019,Siegl2022,Haller2022}. In the RDMFT calculations, we vary the strength of the spin-orbit interaction, $\alpha_{c}$, and the strength of the magnetic field, $B$, in the region according to the results of the classical calculations.

\section{Structure and stability of magnetic skyrmions in the periodic Anderson model}\label{Sec:stability}

To unambiguously identify a magnetic skyrmion, we break the spin translation symmetry of the model in the first DMFT iteration. By this, we select a specific state of the translationally invariant space of ground states. We use two different strategies in our calculations. We either start with a ferromagnetic solution where all $f$ electrons point downwards and flip a single $f$ electron upwards. Alternatively, we directly start with a magnetic skyrmion solution obtained for a different set of parameters. Then, by iterating the DMFT self-consistency equation, we find possible, stable magnetic skyrmion solutions when the algorithm converges.
In the skyrmion phase, both initial states lead to identical DMFT solutions. We show the convergence of the self-energies for a typical skyrmion solution in Appendix~\ref{app:convergence}.

To verify the existence of a magnetic skyrmion, we calculate the spin expectation values of the $c$ and $f$ electrons for each lattice site,
\begin{eqnarray}
\vec S^c_{\vec r}&=&\langle  c^\dagger_{\vec r,\rho_1}\vec \sigma_{\rho_1,\rho_2}c_{\vec r,\rho_2}\rangle,\\
\vec S^f_{\vec r}&=&\langle  f^\dagger_{\vec r,\rho_1}\vec \sigma_{\rho_1,\rho_2}f_{\vec r,\rho_2}\rangle,
\end{eqnarray}
where $\vec r=(i,j)$ corresponds to the coordinates of a lattice site, and $\vec{\sigma}=\left(\sigma^x,\sigma^y,\sigma^z\right)$ is the vector containing the spin space Pauli matrices.
Using these spin expectation values, we calculate the local lattice skyrmion density for the $f$ and $c$ electrons based on the solid angle spanned by three vectors as
\begin{align}
\label{eqnLocalSkyrmionDensity}
    &N^{d}_{\vec r_1,\vec r_2,\vec r_3}=
    \\ & \quad
    \frac{1}{2\pi}\tan^{-1}
    \left(\frac{ \vec S^{d}_{\vec r_1} \cdot(\vec S^{d}_{\vec r_2}\times \vec S^{d}_{\vec r_3})}{\left(\frac{\hbar}{2}\right)^3+\frac{\hbar}{2}\left(\vec S^{d}_{\vec r_1} 
    \vec S^{d}_{\vec r_2}+\vec S^{d}_{\vec r_1} 
    \vec S^{d}_{\vec r_3} + \vec S^{d}_{\vec r_2}
    \vec S^{d}_{\vec r_3}\right)}\right),\nonumber
\end{align}
where $d$ either stands for $f$ or $c$ electrons, $\vec r_1$, $\vec r_2$, and $\vec r_3$ are nearest-neighbor lattice sites spanning an elemental triangle $\langle\vec r_1,\vec r_2,\vec r_3\rangle$ in the densest triangular tessellation of the lattice. The sum of this skyrmion density over all triangles spanning the square lattice yields the skyrmion number
\begin{equation}
    N^{c/f}=\sum_{\langle\vec r_1,\vec r_2,\vec r_3\rangle}N^{c/f}_{\vec r_1,\vec r_2,\vec r_3}.
\end{equation}
Unlike in a classical calculation, the spin expectation values in a quantum model do not need to be $\hbar/2$ in magnitude. In fact, these expectation values are usually smaller due to quantum fluctuations, $\vert \vec S\vert<\hbar/2$. We thus calculate two types of skyrmion densities: One is the quantum skyrmion density/number using unnormalized spin expectation values. The second type is a classical skyrmion density, where we normalize all spin expectation values to $\hbar/2$ before using them in Eq.~(\ref{eqnLocalSkyrmionDensity}). The skyrmion number is an integer when using normalized spin expectation values. When using unnormalized spin expectation values, the skyrmion number is not quantized, and instead, its magnitude is an indicator of the skyrmion stability \cite{Siegl2022}, similar to the scalar chirality defined in Ref.~\cite{Sotnikov2021}.

\begin{figure}
    \centering
    \includegraphics[width=\linewidth]{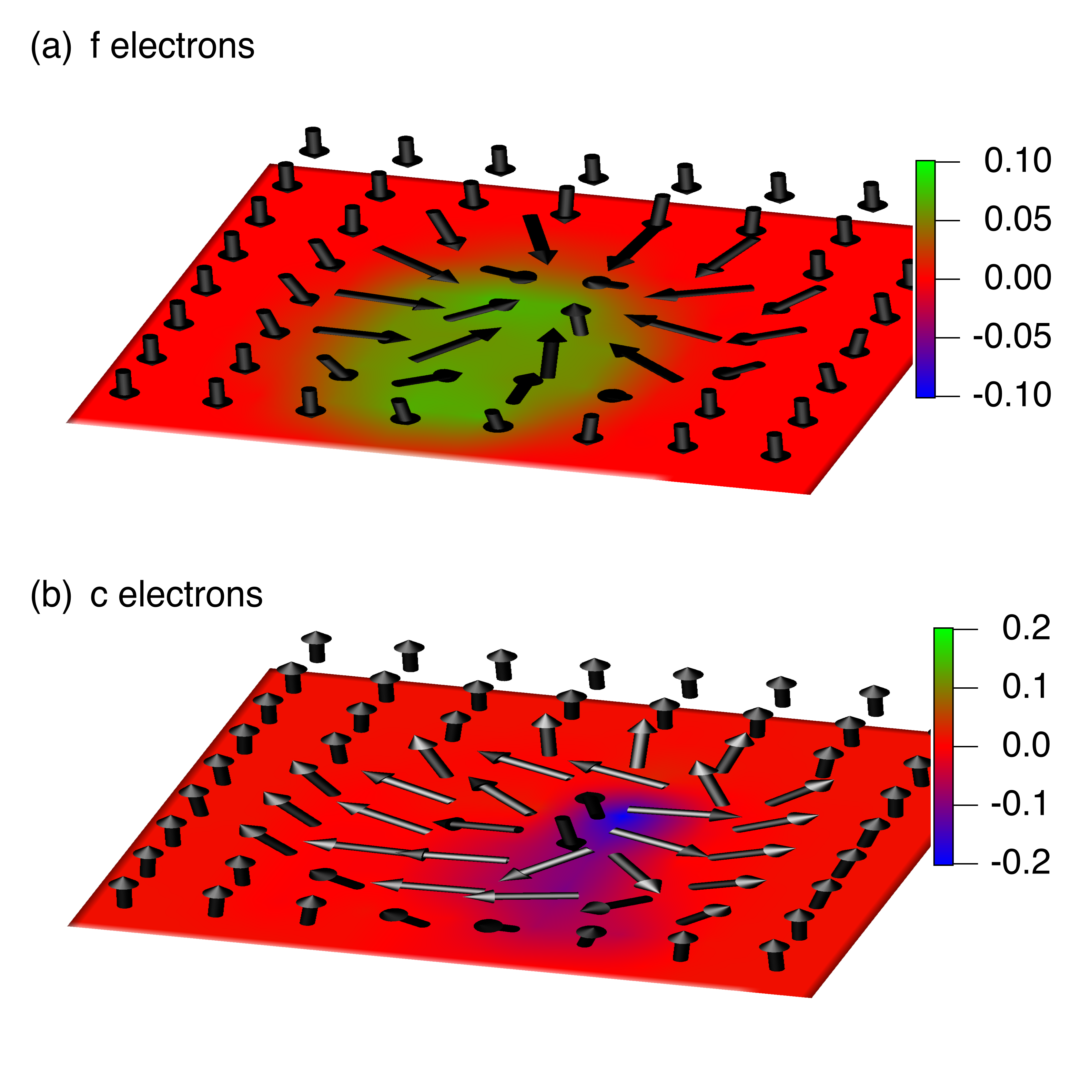}
    \caption{Magnetic skyrmion in an $f$ electron system. A skyrmion forms in the $f$ electrons, depicted by their spin expectation values (a). The color code corresponds to the local skyrmion density in \refEq{eqnLocalSkyrmionDensity}. As a result, an antiskyrmion forms in the itinerant $c$ electrons (b). The antiskyrmion is considerably less polarized, $|\langle S^c \rangle|_a\approx0.03 \frac{\hbar}{2}$. The spin expectation values are shown normalized for better visualization.  Parameters: spin-orbit coupling $\alpha_{c}/t=0.3$ and magnetic field $B/t=0.002$.}
    \label{fig:RepresentativeSkyrmionB02}
\end{figure}

A representative magnetic skyrmion solution is shown in \refFig{fig:RepresentativeSkyrmionB02} calculated for a spin-orbit coupling $\alpha_{c}/t=0.3$ and a magnetic field  $B/t=0.002$. 
We note that within the accuracy of our calculations, we cannot find discernible energy differences between the ferromagnetic configuration and the magnetic skyrmion. The described skyrmions can, therefore, be metastable excitations on a ferromagnetic background with almost vanishing excitation energy or present in the ground state itself. Such an almost degenerate situation is favorable for applications in racetrack devices. If skyrmions were energetically strongly favorable, a skyrmion lattice would form instead of individual skyrmions.
Figure~\ref{fig:RepresentativeSkyrmionB02} shows the spin texture of the $f$ and $c$ electrons underlaid with the local skyrmion density for normalized spin expectation values as 2D color plot, see \refEq{eqnLocalSkyrmionDensity}. 
Due to the local hybridization, $V$, which leads to an effective antiferromagnetic interaction between the $c$ and $f$ electrons, the spins of the $c$ and $f$ electrons mostly point in opposite directions, with deviations in the skyrmion's center, where the Rashba interaction and the itinerant character of the $c$ electrons play a stronger role. 
The combined state corresponds to a bound skyrmion-antiskyrmion pair where the $f$ electrons form a magnetic skyrmion with skyrmion number $N^f=1$ and the $c$ electrons form a magnetic antiskyrmion with skyrmion number $N^c=-1$.
However, in this system, the spin expectation values of the $c$ and $f$ electrons are of very distinct origins and magnitudes.
Because the $f$ electrons are strongly interacting, they form localized magnetic moments, and their spin expectation values in this calculation are approximate $|\langle S^f\rangle|_a\approx0.75 \frac{\hbar}{2}$. They are not perfectly polarized due to the entanglement between the $c$ and the $f$ electrons.
On the other hand, the $c$ electrons are noninteracting, and their spin expectation values vary around $|\langle S^c \rangle|_a\approx0.03 \frac{\hbar}{2}$. The $c$ electrons' polarization is a direct cause of the hybridization with the magnetized $f$ electrons and, thus, a secondary effect.
In this situation, the skyrmion of the $f$ electrons and the antiskyrmion of the $c$ electrons do not annihilate. This is revealed by the finite total skyrmion number calculated with unnormalized spin expectation values. 
This is indeed different from classical skyrmions, where the magnitude of the spin vectors is normalized.
The reduced polarization decreases the topological protection of the magnetic skyrmion. The smaller the spin expectation value, the easier the spin can be flipped, and the magnetic skyrmion is destroyed \cite{Siegl2022}. On the other hand, this facilitates manipulating them as necessary for technical applications.

\begin{figure}
    \centering
\includegraphics[width=\linewidth]{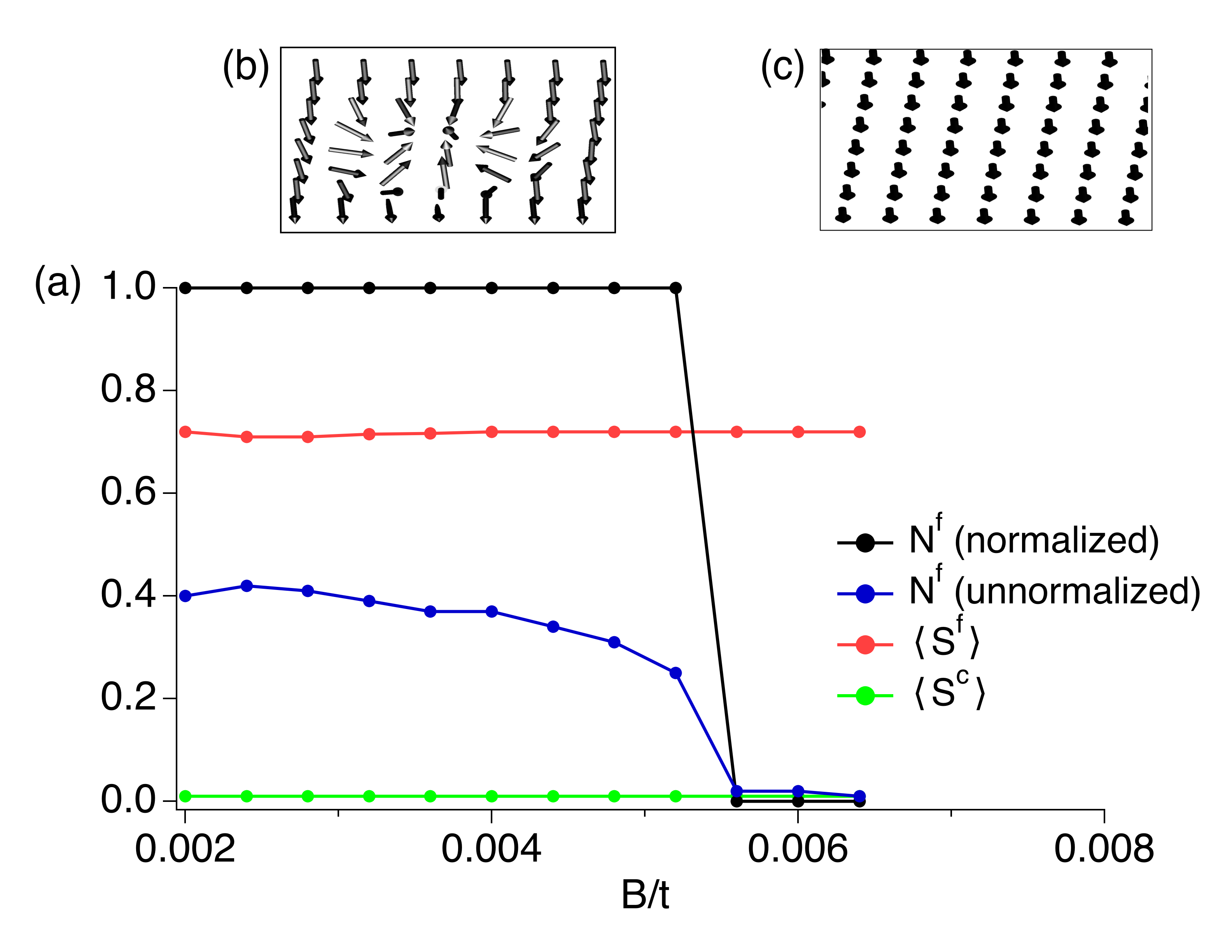}
    \caption{Magnetic field dependence of magnetic skyrmions for spin-orbit coupling $\alpha_c=0.3t$. Panel (a) shows the normalized and unnormalized skyrmion number and average spin expectation values of the $c$ and $f$ electrons.    
    The skyrmion number drops to zero at $B/t\approx 0.0056$, and the spins align ferromagnetically, consistent with studies on quantum skyrmions in spin lattices \cite{Sotnikov2021,Haller2022}. 
    The average spin expectation value $|\langle S^{f/c}\rangle|_a$ of the $c$ and $f$ electrons alone does not indicate this phase transition. Panels (b) and (c) show typical $f$-spin configurations for small (skyrmionic configuration at $B/t=0.002$) and large magnetic fields (ferromagnetic configuration at $B/t=0.006$), respectively.}
    \label{fig:Bdependence}
\end{figure}

Next, we analyze the stability of the magnetic skyrmion for different magnetic field strengths, as shown in Fig.~\ref{fig:Bdependence}.
As stated above, we generally apply a small magnetic field which helps to stabilize the magnetic skyrmion against spin spiral solutions \cite{Sotnikov2021,Haller2022}.
In Fig.~\ref{fig:Bdependence}(a), we show the skyrmion number using normalized and unnormalized spins, respectively. We observe that, while the skyrmion number (normalized) is constantly one for $B/t \lesssim 0.0056 = B_c$, the skyrmion number using unnormalized spin expectation values is $N^f\approx 0.4$ and gradually drops for an increased magnetic field until $B_c$ is reached. The difference between these numbers demonstrates the relevance of quantum effects to the system at hand.
We furthermore show the average spin expectation values of the $c$ and $f$ electrons, indicating that the $f$ electrons are considerably more stronger polarized than the $c$ electrons. Furthermore, for magnetic fields stronger than $B_c$, we only find ferromagnetic solutions. Figures~\ref{fig:Bdependence}(b) and (c) give representative spin textures of the $f$ electrons for the corresponding parameter regimes, i.e., small and large magnetic fields. 

\begin{figure}
    \centering
    \includegraphics[width=\linewidth]{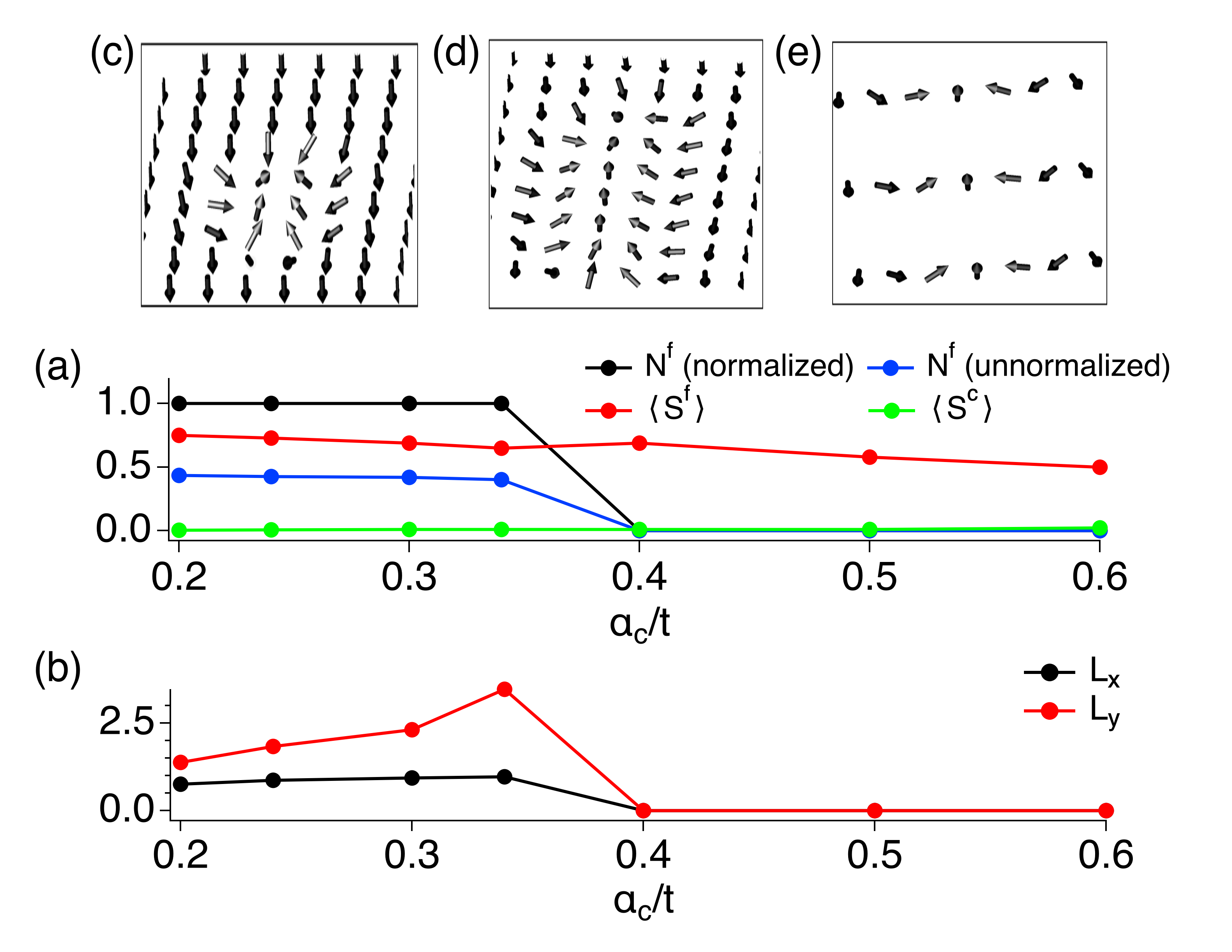}
    \caption{Dependence of magnetic skyrmions on the spin-orbit coupling for $B/t = 0.002$. Panel (a) shows the skyrmion number (normalized), skyrmion number (unnormalized), and averaged spin expectation values ($c$- and $f$-electrons), $\langle \vec S^f\rangle_a$ and  $\langle \vec S^c\rangle_a$, for different strengths of the Rashba interaction. The skyrmion changes to a spin density wave at $\alpha_c/t=0.4$. Panel (b) shows the extension of the skyrmion in the $x$ and $y$ direction; see \refEq{eqnSizeOfSkyrmion}. Panels (c)-(e) show representative spin configurations for small ($\alpha_c=0.2t$), medium  ($\alpha_c=0.35t$), and large  ($\alpha_c=0.5t$) spin-orbit interaction. }
    \label{fig:PhaseV}
\end{figure}

We next analyze the structure of the skyrmion depending on the strength of the Rashba spin-orbit coupling $\alpha_{c}$.
We show the skyrmion number using normalized spin expectation values, the skyrmion number using unnormalized spin expectation values, and the average spin expectation values ($\langle S^f\rangle$ and $\langle S^c\rangle$) in Fig.~\ref{fig:PhaseV}(a).
Increasing the Rashba interaction, the $f$-electron spin expectation value is slightly suppressed, while the $c$-electron spin expectation value slightly increases. This increase in the $c$ electron magnetization can be explained by the stronger coupling between the $c$ and $f$ electrons.
While we need $\alpha_{c}/t>0$ to create a finite DM interaction that stabilizes the magnetic skyrmion, we see that for increasing $\alpha_{c}$ the skyrmion gets destabilized and for $\alpha_{c}/t>0.4$, magnetic skyrmions become unstable indicated by the vanishing skyrmion number.
To analyze this transition further, we calculate the average size of the skyrmion. First, the center of the skyrmion created by the $f$ electrons is \begin{equation}
\vec R_S=\sum_{\langle \vec r_1,\vec r_2,\vec r_3 \rangle} N^f_{\vec r_1,\vec r_2,\vec r_3}\frac{\vec r_1+\vec r_2+\vec r_3}{3},
\label{eq:center}
\end{equation}
where $\vec r_1$, $\vec r_2$, and $\vec r_3$ are the coordinates of the lattice sites spanning the elemental triangle as explained below \refEq{eqnLocalSkyrmionDensity}.
The extension of the skyrmion in the $x$ and the $y$-direction $\vec L=(L_x,L_y)$ is then given as
\begin{eqnarray}
\label{eqnSizeOfSkyrmion}
\vec L_x^2&=&\sum_{\vec r_1,\vec r_2,\vec r_3} N^f_{\vec r_1,\vec r_2,\vec r_3}\left(\frac{\vec x_1+\vec x_2+\vec x_3}{3}-\vec x_S\right)^2,\\
\vec L_y^2&=&\sum_{\vec r_1,\vec r_2,\vec r_3} N^f_{\vec r_1,\vec r_2,\vec r_3}\left(\frac{\vec y_1+\vec y_2+\vec y_3}{3}-\vec y_S\right)^2,
\end{eqnarray}
where $x_i$ ($y_i$) is the $x$ ($y$) component of the position $\vec r_i$ and the center of the skyrmion is $\vec R_s=(x_s,y_s)$.
In Fig.~\ref{fig:PhaseV}(b), we show the extension of the skyrmion in the $x$ and $y$-direction depending on the Rashba spin-orbit interaction. We see that while the average extension of the skyrmion in the $x$ direction remains unchanged when increasing $\alpha_{c}$, the magnetic skyrmion is strongly elongated in the $y$ direction. At $\alpha_{c}/t\approx 0.4$, the magnetic skyrmion changes into a spiral phase, again consistent with findings for quantum skyrmions on nonelectronic spin lattices \cite{Sotnikov2021,Haller2022}.
Representative spin textures of the $f$ electrons are shown in Fig.~\ref{fig:PhaseV}(c)-(e), depicting a skyrmion for small $\alpha_{c}$ (c), an elongated skyrmion close to the phase transition (d), and a spiral wave for large $\alpha_{c}$ (e).
For larger values of the spin-orbit coupling, we do not find stable skyrmion solutions.

Finally, we note that we have confirmed the stability of the quantum skyrmion phase for smaller lattice sizes, such as 7x7. Magnetic skyrmions remain stable as long as their elongation is smaller than the lattice width. Furthermore, we do not find an even/odd effect in the lattice width, which can be understood by the fact that all spins are ferromagnetically aligned far away from the magnetic skyrmion, irrespective of changes of the lattice sizes once it exceeds the size of the quantum skyrmion.

\section{Charge-driven quantum skyrmions --- the onset of the quantum skyrmion Hall effect}\label{Sec:SkyrmionHall}
Finally, we study the response of the identified stable skyrmion textures to an applied charge current in the itinerant $c$ electrons.
To do this, we calculate the change in the spin expectation values of all lattice sites in linear response theory. 
We focus on describing the time-transient behavior of the system. A description of the nonequilibrium steady state poses considerable numerical challenges, as discussed in the concluding remarks. 

In linear response, the change in an expectation value of operator $A$ resulting from a perturbation $B$ is given by
\begin{eqnarray}
\langle A\rangle(\tau)&=&\langle A\rangle(0)+\! \int_0^\tau \! d\tau^\prime\, X_{AB}(\tau-\tau^\prime)\\ 
X_{AB}(\tau-\tau^\prime)&=&i\Theta(\tau-\tau^\prime)\langle [A(\tau),B(\tau^\prime)]\rangle,
\label{eq:linearresponse}
\end{eqnarray}
where $\Theta(\tau)$ is the Heaviside step function. 
Because we are interested in the linear response of the spin expectation values of the $f$ electrons to a charge current in the itinerant electrons, we use
\begin{eqnarray}
A&=&f^\dagger_{\vec r,\rho_1}\sigma^{x/y/z}_{\rho_1\rho_2}f_{\vec r,\rho2} = S^{x/y/z},\\
B&=&J^c=-i J \sum_{i,j,\sigma}\left(c^\dagger_{i+1,j,\sigma}c_{i,j,\sigma}-c^\dagger_{i-1,j,\sigma}c_{i,j,\sigma}\right),
\end{eqnarray}
where $A$ corresponds exactly to the local spin of an $f$ electron, and $B$ is the charge current operator in the $c$ electrons. 
\begin{figure}
    \centering
    \includegraphics[width=\linewidth]{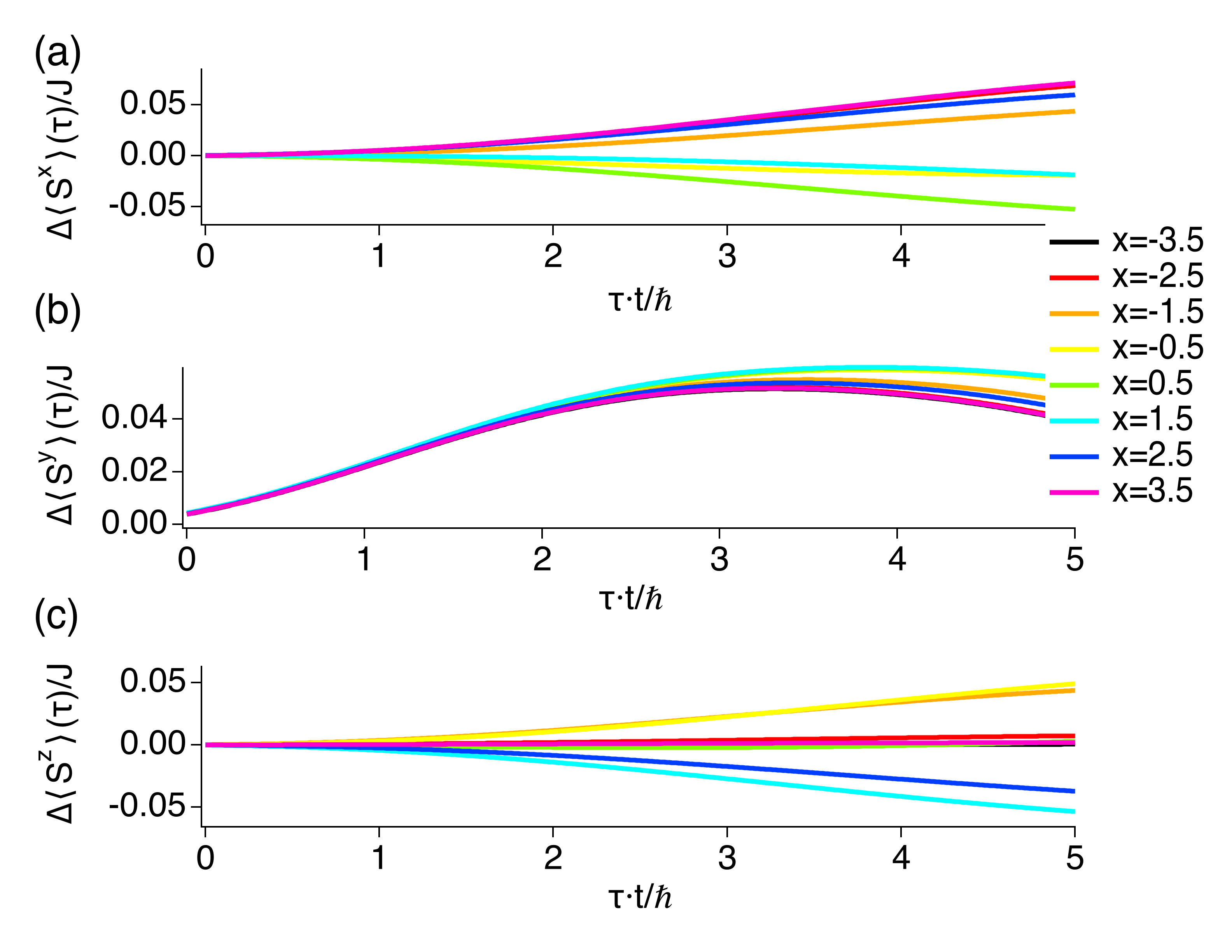}    
       \caption{Time-resolved change of the spin expectation values $\Delta\langle S^x \rangle(\tau)$ (a), $\Delta\langle S^y \rangle(\tau)$ (b), and $\Delta\langle S^z \rangle(\tau)$ (c) scanned in the $x$ direction across the center of the magnetic skyrmion at lattice sites $(x_S+x,y_S)$, calculated by linear response theory for $\alpha_{c}/t=0.3$. Here, $\vec{R}_S=(x_S,y_S)$ is the center of the skyrmion, see Eq.~(\ref{eq:center}). The change of the spin expectation values is normalized by the strength of the current, $J$. The expectation values start oscillating when the validity regime of the linear response theory is left. \label{fig:Skt}}
\end{figure}

\begin{figure}
    \centering
    \includegraphics[width=0.48\linewidth]{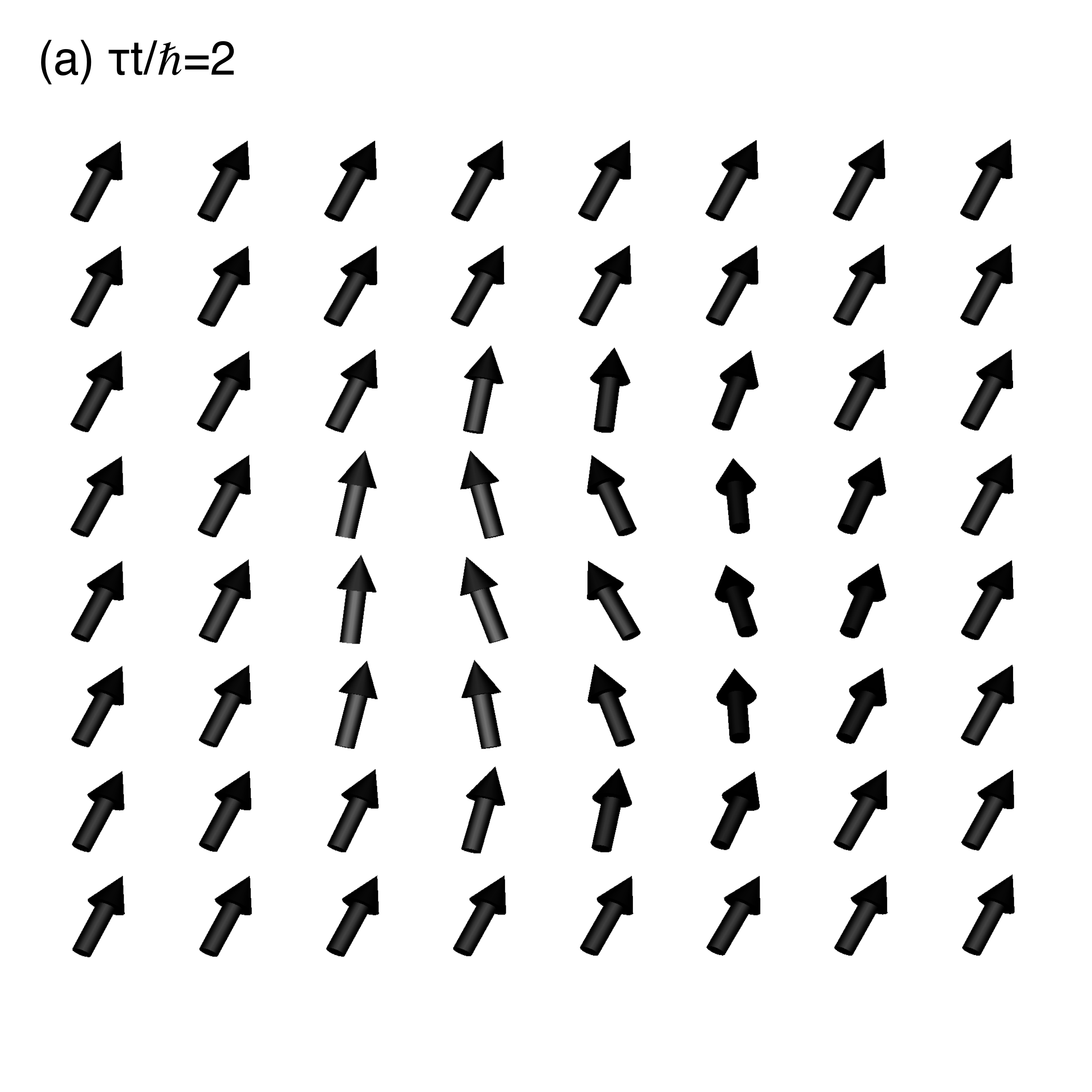}  
     \includegraphics[width=0.48\linewidth]{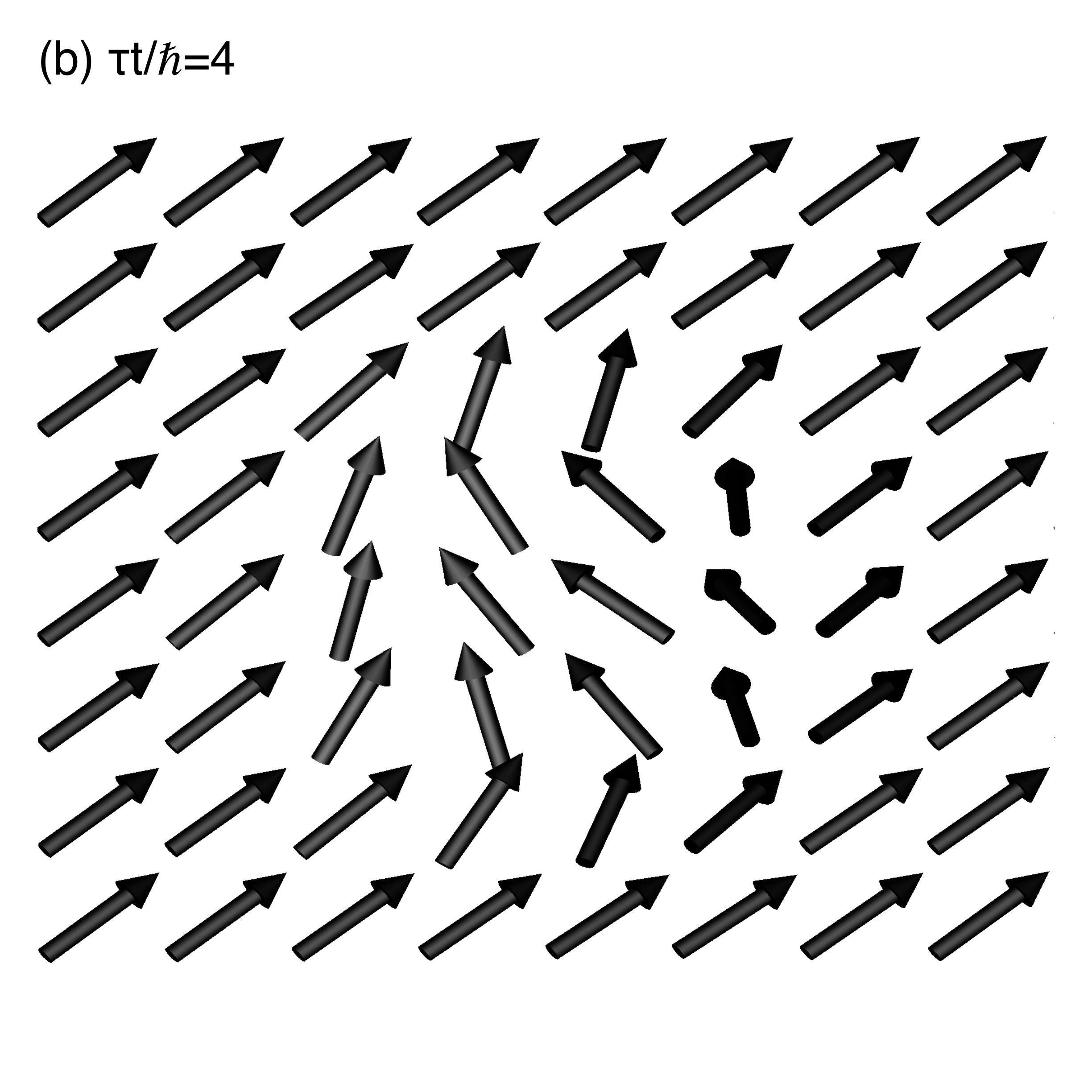}  
     \caption{Changes in the spin expectation values at (a) $\tau \cdot t/\hbar=2$ and (b) $\tau \cdot t/\hbar=4$ for $\alpha_{c}/t=0.3$. 
     The length of the vectors has been multiplied by five in both figures to enhance visibility.
     The actual magnitude of the changes are (a) $|\Delta \left\langle\vec S\right\rangle(\tau)| \approx 0.04 \frac{\hbar}{2}J$ and (b) $|\Delta \left\langle\vec S\right\rangle(\tau)| \approx 0.07 \frac{\hbar}{2}J$.\label{fig:Skt_arrows}}
\end{figure}

\begin{figure}
    \centering
    \includegraphics[width=\linewidth]{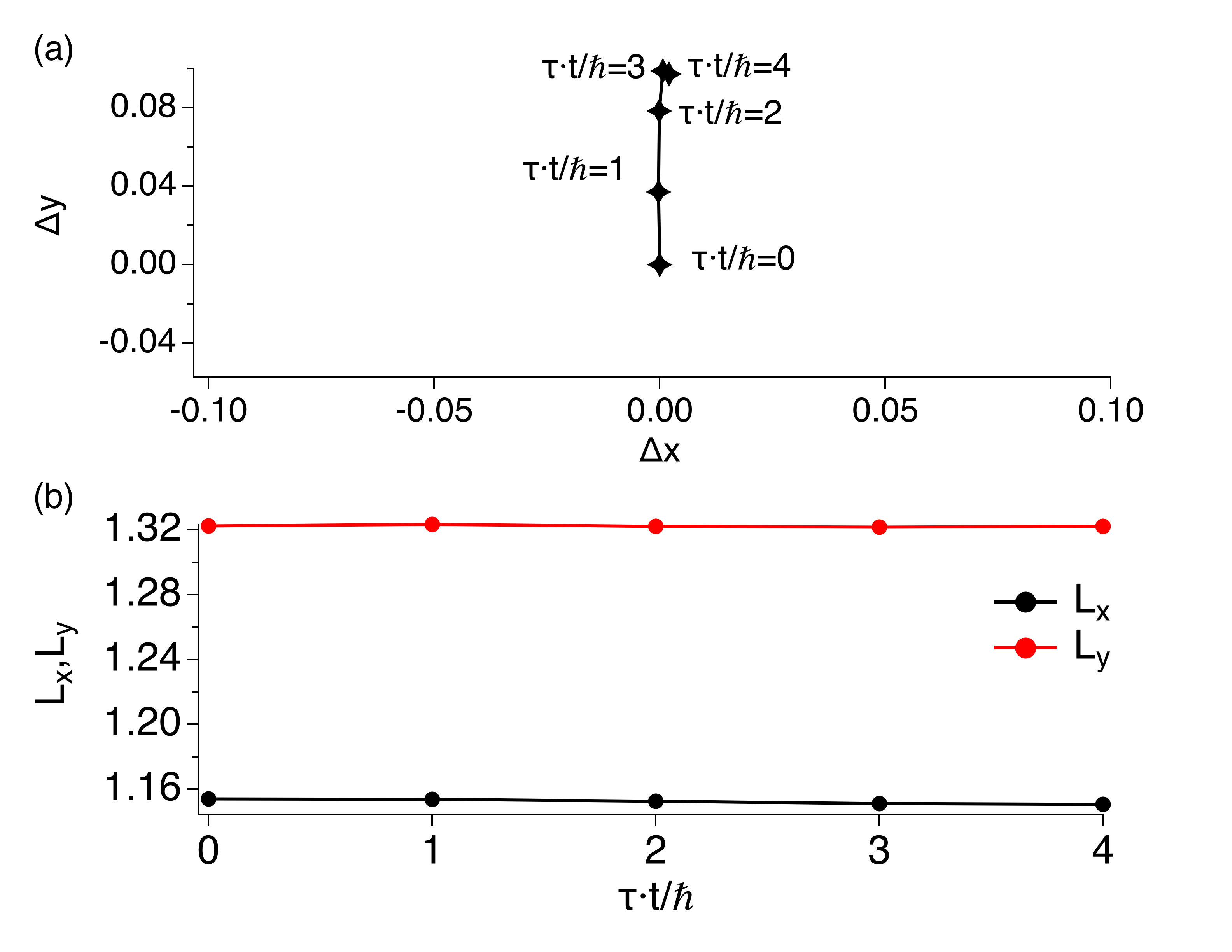}
    \caption{Onset of the quantum skyrmion Hall effect for $\alpha_c/t=0.3$ and $B/t=0.002$: Shown are the center (a) and the size (b) of the quantum magnetic skyrmion depending on time, calculated by linear response theory. The quantum skyrmion starts moving almost perpendicularly to the applied current, indicating a Magnus angle close to $90\deg$.
    When the validity of the linear response calculations is left, the skyrmion slows down. The size of the skyrmion stays constant over time, indicating a negligible smearing of the structure compared to its motion.
    }
    \label{fig:Skcenter_t}
\end{figure}
For these operators, Eq.~(\ref{eq:linearresponse}) corresponds to a nonlocal two-particle Green's function. Using the DMFT approximation, where vertex corrections in nonlocal Green's functions vanish, we write these two-particle Green's functions as the convolution of two single-particle Green's functions.
Then, we calculate the time evolution of the spin expectation values of all spins. Because the self-energy depends on the lattice site, also the time evolution of the spin expectation value depends on the lattice site. This is shown in Fig.~\ref{fig:Skt}, where we show the change of the $x$, $y$, and $z$ component of the spin expectation values  $\Delta\langle S^x\rangle (\tau)$, $\Delta\langle S^y\rangle (\tau)$, and $\Delta\langle S^z\rangle (\tau)$ along the $x$-direction of the lattice across the center of the skyrmion solution for $\alpha_{c}/t=0.3$ (shown in Fig.~\ref{fig:RepresentativeSkyrmionB02}). 
Specifically, the spin expectation values are shown for lattice sites $(x_S+x,y_S)$, where $\vec{R}_S=(x_S,y_S)$ is the center of the skyrmion, see \refEq{eq:center}.
$\Delta\langle S_x\rangle$ and $\Delta\langle S_z\rangle$ show a strong dependence on the position close to the center of the skyrmion. $\Delta\langle S_z\rangle$ changes even its sign when changing the position from the left of the center to the right of the center. On the other hand, $\Delta\langle S_y\rangle$ is nearly independent of the lattice site. Also, while $\Delta\langle S_z\rangle$ becomes small for spins far away from the skyrmion center, $\Delta\langle S_x\rangle$ and $\Delta\langle S_y\rangle$ are nonzero. Thus, even in the ferromagnetic region away from the skyrmion center, $\langle S_x\rangle$ and $\langle S_y\rangle$ change.
This rotation of the spin in the ferromagnetic state when a charge current is applied is explained by the Edelstein and the magnetoelectric effect \cite{PhysRevB.97.115128}; in a system where the Fermi surface is split due to the Rashba spin-orbit coupling, a charge current results in an accumulation of spin. This can be seen here as a rotation of the spin expectation values in the $x$ and the $y$ direction, even far away from the magnetic skyrmion. 
 Furthermore, we emphasize that the linear-response results only remain valid within sufficiently small times $\tau$. In \refFig{fig:Skt}, we see that the initial linear trend in $\tau$ is reduced, and, as an expected artifact from linear response theory,  all spin expectation values start oscillating after a certain individual time. The change of all spin expectation values for $\tau\cdot t/\hbar=2$ and $\tau\cdot t/\hbar=4$  are visualized in Fig.~\ref{fig:Skt_arrows} as arrows. Each arrow corresponds 
 to the direction of the local change in the spin expectation values $\Delta\langle \vec{S} \rangle (\tau)$.
  The actual length of each change is $|\Delta \left\langle\vec S\right\rangle(\tau)| \approx 0.04 \frac{\hbar}{2}J$ and $|\Delta \left\langle\vec S\right\rangle(\tau)| \approx 0.07 \frac{\hbar}{2}J$ for $\tau\cdot t/\hbar=2$ and $\tau\cdot t/\hbar=4$,respectively.
 We see that in the ferromagnetic region, all spins are rotated in the same direction. Only close to the magnetic skyrmion, 
    the change in the spin expectation values significantly depends on the lattice site.
    
Finally, we take the time evolution of each spin on the lattice and calculate the skyrmion density and the time-dependent size and position of the skyrmion according to Eqs.~(\ref{eq:center}-\ref{eqnSizeOfSkyrmion}). By \refEq{eq:linearresponse}, we find that the center of the skyrmion moves almost perpendicularly to the applied current, as shown in Fig.~\ref{fig:Skcenter_t}(a). While the current is applied in the $x$ direction, the skyrmion moves in the positive $y$ direction. 
Thus, our results demonstrate the onset of a quantum skyrmion Hall effect with a Magnus angle close to $90$ degrees.
Notably, the size of the skyrmion effectively remains constant during the motion, shown in \refFig{fig:Skcenter_t}(b).

\section{Discussion}\label{Sec:discussion}
In conclusion, we show that noncentrosymmetric $f$-electron systems with spin-orbit coupling in the presence of a small external magnetic field can host nano quantum skyrmions in the ground state, and we demonstrate the onset of the quantum skyrmion Hall effect upon applying a charge current, which is accompanied by an Edelstein and magnetoelectric effect.

The reason for the stability of the quantum skyrmion is an effective DM interaction generated by the spin-orbit interaction and a local density-density interaction.
Despite the itinerant $c$ electrons being magnetized like an antiskyrmion, the quantum skyrmions of the $f$ electrons remain stable and dominate the physical behavior of the system because of its considerably stronger polarization due to strong correlations.
Concerning the quantum skyrmion Hall effect, we observe a Magnus angle close to $90\deg$. This is consistent with the behavior of classical skyrmions, where the Magnus angle increases when the size of the skyrmions is smaller or when dissipative effects are small  \cite{Litzius2016,Jiang2016DirectObservationSkyrmionHallEffect}. Both is the case for the observed nano quantum skyrmions. Furthermore, no quantum skyrmion pinning is visible in our study.
We note that our method can only describe the onset of the skyrmion motion. In particular, in a full nonequilibrium calculation, time-dependent spin expectation values would lead to time-dependent self-energies.
The system would adapt to the changed spin expectation values and backaction effects would alter our conclusions when the linear-response regime is left. For example, linear response theory can permanently decrease the polarization locally, ultimately resulting in a site with vanishing spin polarization. However, this situation is energetically unfavorable due to the strong density-density interaction. Thus, in a full nonequilibrium calculation, self-energies will change in a way that an atom with vanishing spin polarization is prevented, rendering the quantum magnetic skyrmion stable and letting it continue its motion perpendicular to an applied current.
Yet, a full nonequilibrium calculation, as well as a steady-state analysis, goes beyond the scope of the current paper and is left for future work.

We note that other forms of spin-orbit interaction also lead to stable nano quantum skyrmions in the $f$-electron system at hand. 
We show the results for a different form of the spin-orbit interaction, where the momenta couple to the same spin direction, in Appendix~\ref{app:diff_SOI}.
Also in these systems, the spin-orbit interaction results in a spin accumulation when a current is applied, which leads to a site-dependent change of the spin expectation values, and to a skyrmion Hall effect.  These results emphasize that the existence of magnetic skyrmions in strongly correlated $f$-electron systems with spin-orbit coupling and the skyrmion Hall effect is a general effect.

\begin{acknowledgments}
All authors acknowledge funding by the Kyoto University - Hamburg University (KU-UHH) international partnership funding program for 2021 and 2022.
R.P. is supported by JSPS KAKENHI No.~JP18K03511 and JP23K03300. 
Parts of the numerical simulations in this work have been done using the facilities of the Supercomputer Center at the
Institute for Solid State Physics, the University of Tokyo.
J. N.-S. acknowledges support by the Cluster of Excellence ``CUI: Advanced Imaging of Matter'' of the Deutsche Forschungsgemeinschaft (DFG) – EXC 2056 – project ID 390715994 and the Universit{\"a}t Hamburg’s Next Generation
Partnership funded under the Excellence Strategy of the
Federal Government and the L{\"a}nder. 
T. P. acknowledges funding by the DFG (project no. 420120155) and the European Union (ERC, QUANTWIST, project number 101039098). Views and opinions expressed are however those of the authors only and do not necessarily reflect those of the European Union or the European Research Council. Neither the European Union nor the granting authority can be held responsible for them. 
\end{acknowledgments}

\appendix
\section{Different form of spin-orbit interaction}
\label{app:diff_SOI}

\begin{figure}
    \centering
    \includegraphics[width=\linewidth]{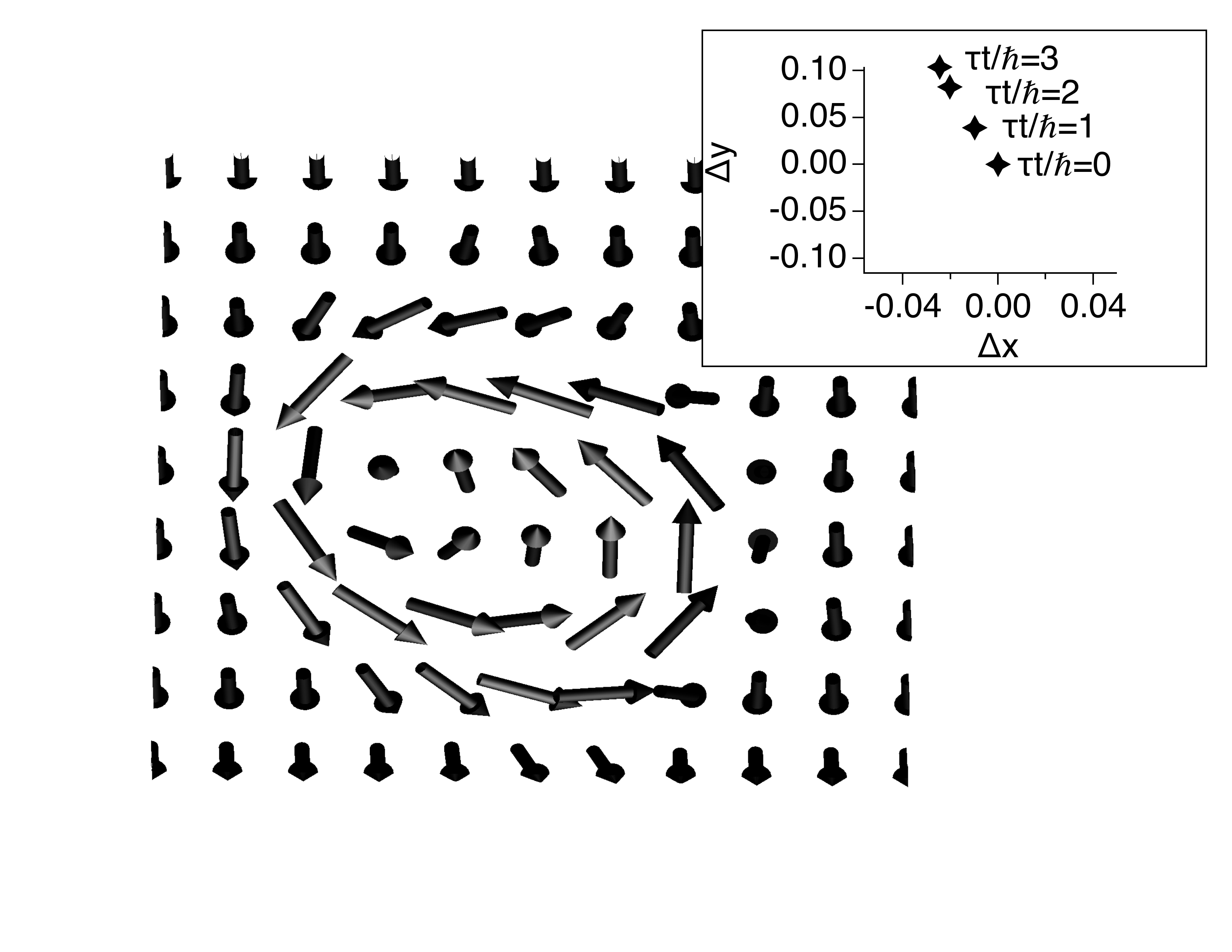}
        \includegraphics[width=\linewidth]{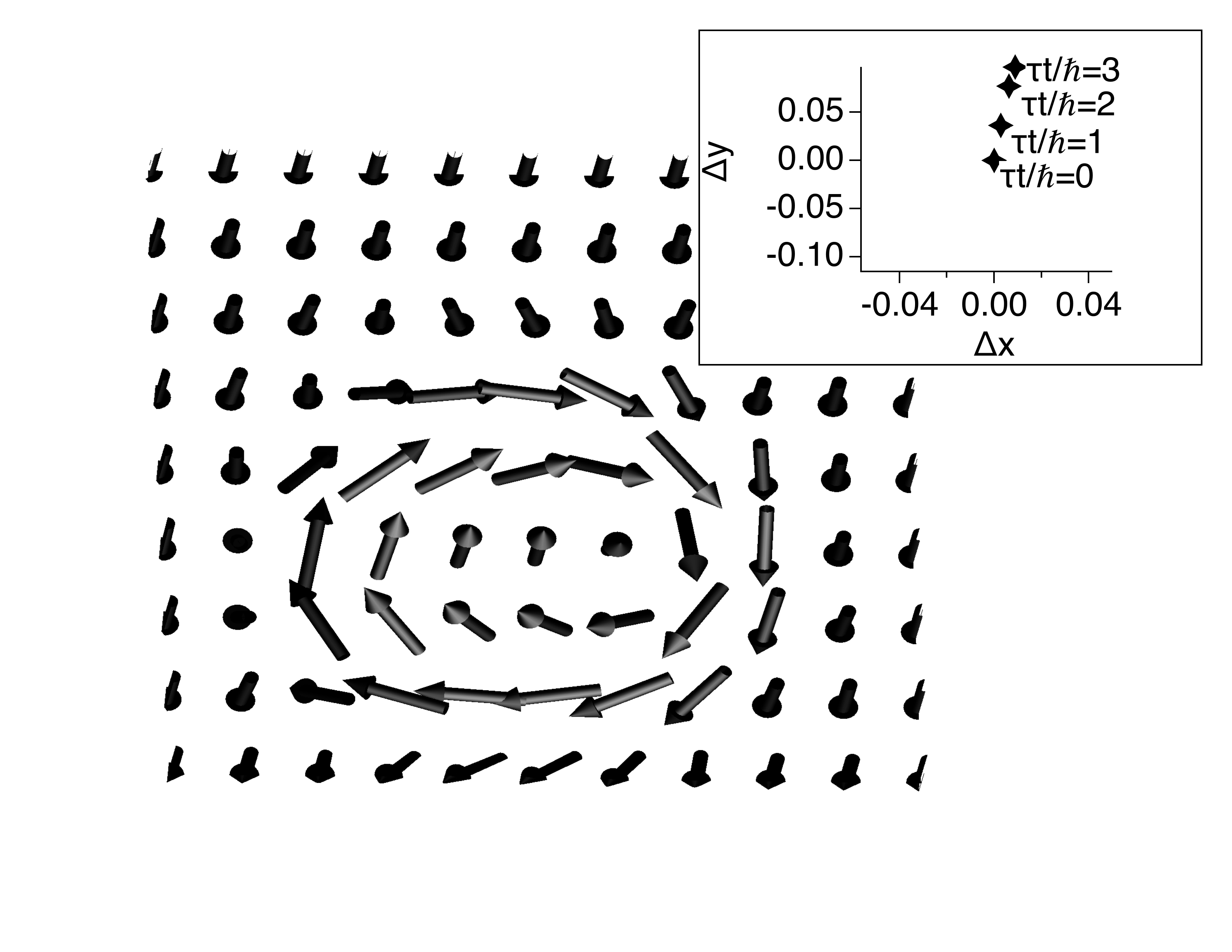}
    \caption{Spin texture including a magnetic skyrmion for $\alpha_c/t=0.3$ (top panel) and $\alpha_c/t=-0.3$ (bottom panel). The insets show the initial motion of the center of the skyrmion after a charge current in $x$ direction is applied.}
    \label{fig:Skcenter_t_app}
\end{figure}

To demonstrate that our results are robust for different types of spin-orbit coupling, we repeat our analysis using a spin-orbit interaction of the form
\begin{eqnarray}
H_{SOI}(\vec{k})&=& 
 2\alpha_c( \sin(k_x)  \vec{c}^\dagger_{\vec k} \sigma^x \tau^x  \vec{c}_{\vec k} 
 \nonumber \\ 
 &+& 
\sin(k_ y) \vec{c}^\dagger_{\vec k} \sigma^y \tau^x \vec{c}_{\vec k}).
\end{eqnarray}
The rest of the Hamiltonian, including the two-particle interaction, is unchanged compared to the main text.
In Fig.~\ref{fig:Skcenter_t_app}, we show two RDMFT solutions, including magnetic skyrmions, for $\alpha_c=\pm 0.3 t$, where
we again use a small magnetic field, $B/t=0.002$, to stabilize the magnetic skyrmion 
\cite{Sotnikov2021,Lohani2019,Siegl2022,Haller2022}.
The change in the sign of the spin-orbit interaction leads to a change in the rotation direction of the spin texture.

Furthermore, we apply a charge current in the $x$ direction for both solutions and find that the center of the quantum magnetic skyrmion dominantly moves into the positive $y$-direction. This is explained as follows: 
The change in the sign of the spin-orbit interaction leads not only to a reversal of the spin rotation inside the skyrmion but also changes the sign of the Edelstein and magnetoelectric effect. Thus, spins in these two examples are rotated in opposite directions when current is applied. As a result, both magnetic skyrmions move into the same, the positive $y$ direction.

\section{Convergence of the real-space DMFT}
\label{app:convergence}

\begin{figure}
    \centering
    \includegraphics[width=\linewidth]{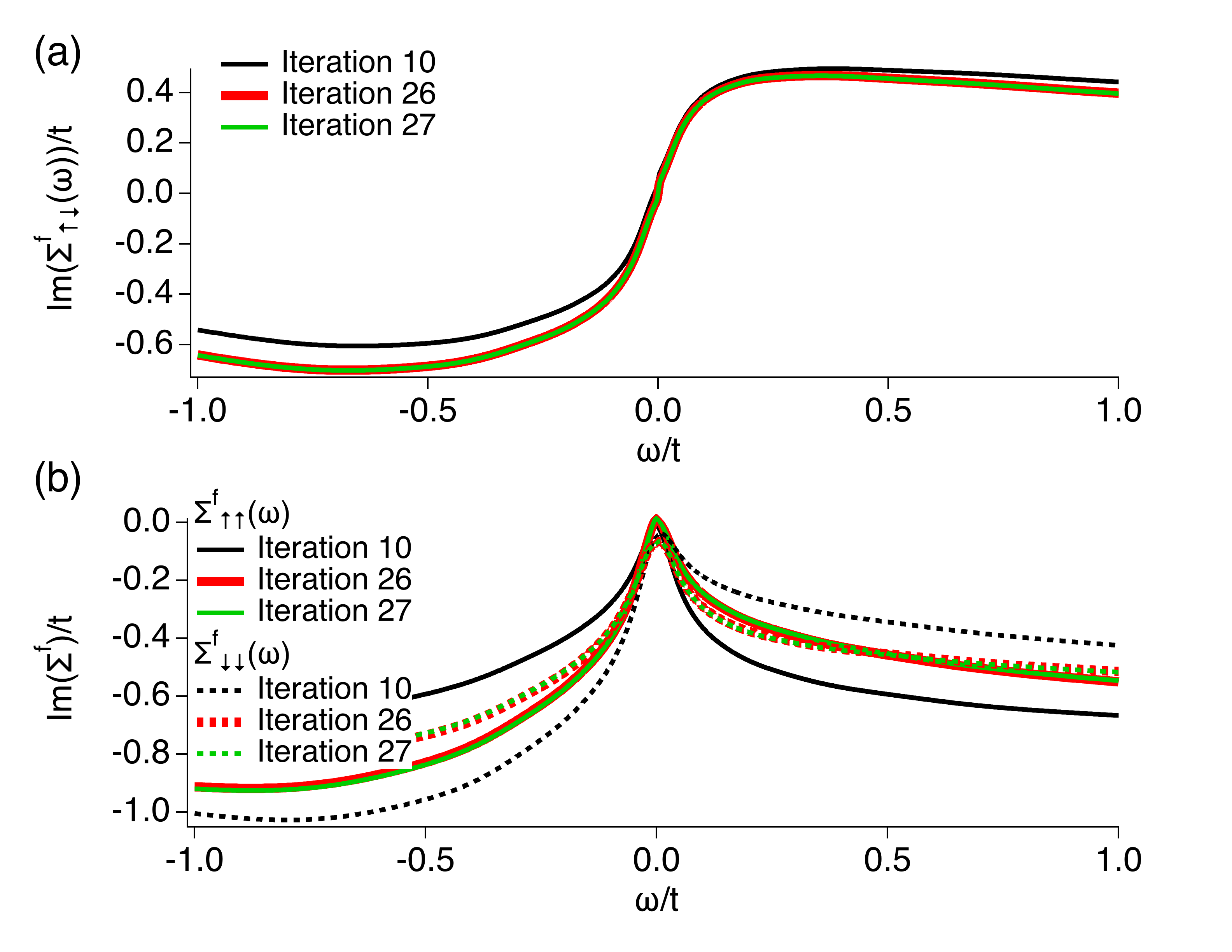}
       \caption{Convergence of the diagonal and off-diagonal self-energies of the $f$ electrons for a lattice site left of the skyrmion center.}
    \label{fig:self_energy33}
\end{figure}
\begin{figure}
    \centering
    \includegraphics[width=\linewidth]{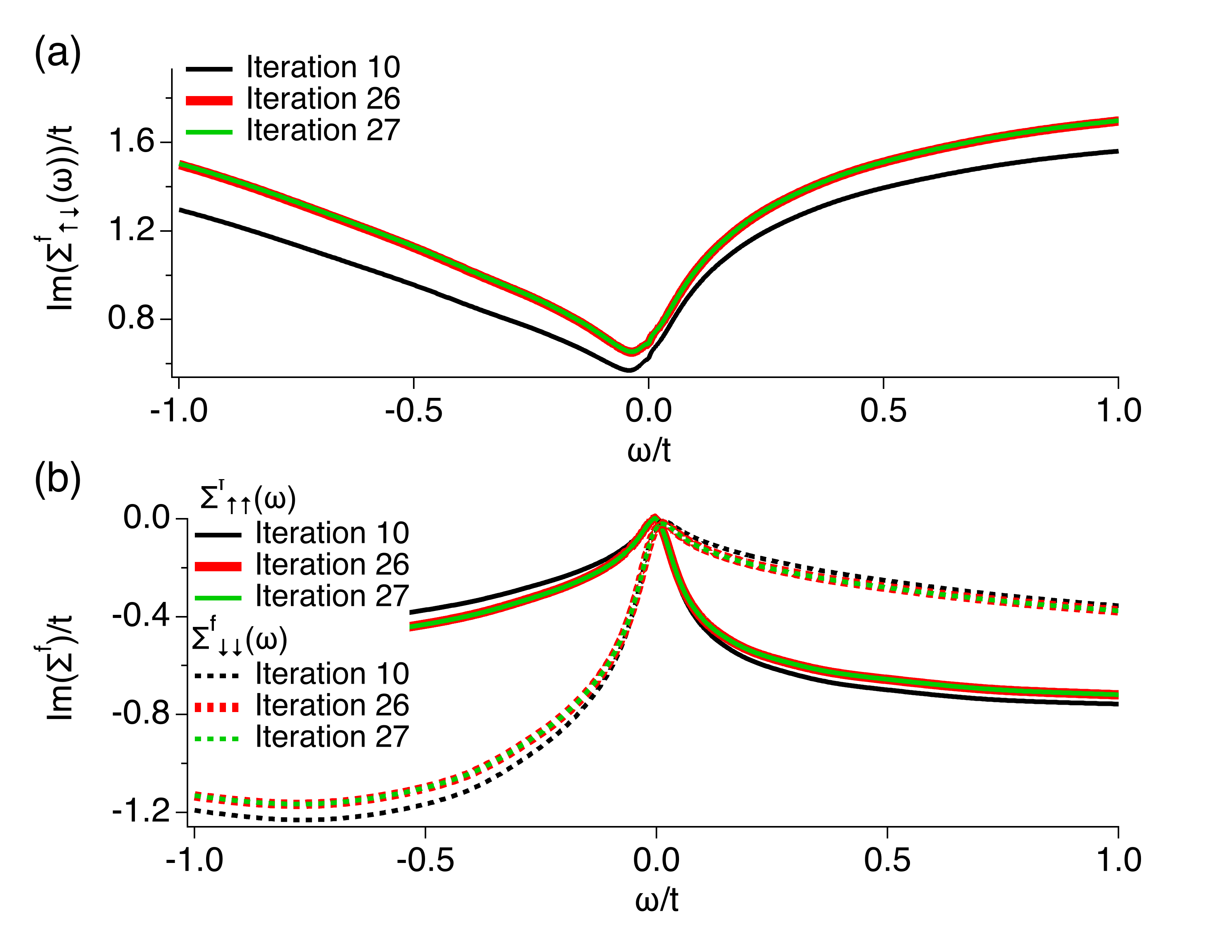}
       \caption{Convergence of the diagonal and off-diagonal self-energies of the $f$ electrons for a lattice site below the skyrmion center.}
    \label{fig:self_energy41}
\end{figure}
In this appendix, we demonstrate the convergence of the real-space DMFT in the magnetic skyrmion phase for $\alpha_c/t=0.3$. In Figs.~\ref{fig:self_energy33} and \ref{fig:self_energy41}, we show representative self-energies of the $f$ electrons for two different lattice sites and different DMFT iterations. Panel (a) shows the off-diagonal self-energy, $\Sigma_{\uparrow\downarrow}(\omega)$, and panel (b) shows the diagonal self-energies, $\Sigma_{\uparrow\uparrow}(\omega)$ and $\Sigma_{\downarrow\downarrow}(\omega)$. On average, we need $20$-$30$ DMFT iterations (depending on the parameters) to obtain a converged magnetic skyrmion solution. In both figures, we see that while the self-energy of the 10th iteration qualitatively shows the same behavior as iterations 26 and 27, quantitatively, it still differs from the converged self-energy. On the other hand, the self-energies of the 26th and 27th iterations lie on top of each other.


\bibliography{libraryShort}

\end{document}